\documentclass[11pt,letterpaper,english,american]{aastex6}
\usepackage[T1]{fontenc}
\usepackage[latin9]{luainputenc}
\usepackage{array}
\usepackage{float}
\usepackage{multirow}
\usepackage{amssymb}
\usepackage{graphicx}

\makeatletter

\@ifundefined{pageheight}{\let\pageheight\pdfpageheight}{}
\@ifundefined{pagewidth}{\let\pagewidth\pdfpagewidth}{}
\pageheight\paperheight
\pagewidth\paperwidth

\providecommand{\tabularnewline}{\\}

\usepackage{lineno}

\makeatother

\usepackage{babel}
\begin{document}
\selectlanguage{english}%
\shortauthors{Rufu \& Canup}

\shorttitle{Triton's Evolution within a Satellite System}
\selectlanguage{american}%

\title{Triton's Evolution with a Primordial Neptunian Satellite System}

\author{Raluca Rufu }

\affil{Department of Earth and Planetary Sciences, Weizmann Institute of
Science, Rehovot 76100, Israel.}

\author{Robin M. Canup $^*$}

\altaffiliation{$^*$ Planetary Science Directorate}

\affil{Southwest Research Institute, Boulder, Colorado 80302, USA.}

\email {raluca.rufu@weizman.ac.il}
\begin{abstract}
\noindent The Neptunian satellite system is unusual. The major satellites
of Jupiter, Saturn and Uranus are all in prograde, low inclination
orbits. Neptune on the other hand, has the fewest satellites and most
of the system's mass is within one irregular satellite, Triton. Triton
was most likely captured by Neptune and destroyed the primordial regular
satellite system. We investigate the interactions between a newly
captured Triton and a prior Neptunian satellite system. We find that
a prior satellite system with a mass ratio similar to the Uranian
system or smaller has a substantial likelihood of reproducing the
current Neptunian system, while a more massive system has a low probability
of leading to the current configuration. Moreover, Triton's interaction
with a prior satellite system may offer a mechanism to decrease its
high initial semimajor axis fast enough to preserve small irregular
satellites (Nereid-like), that might otherwise be lost during a prolonged
Triton circularization via tides alone.
\end{abstract}

\keywords{planets and satellites: dynamical evolution and stability}

\section{Introduction}

Models of giant planet gas accretion and satellite formation suggest
that gas giants may typically have prograde regular satellite systems
formed within circumplanetary gas disks produced during their gas
accretion, consistent with the general properties of the satellite
systems of Jupiter and Saturn (e.g., \citealp{stevenson1986,Lubow1999,Canup2002,MOSQUEIRA2003,Sasaki2010,Ogiara2012}).
The origin of Uranus' satellites remains unclear, as they may have
similarly formed as a result of Uranus' limited gas accretion (e.g.,
\citealp{Pollack1991,MOSQUEIRA2003,Canup_2006}), or as a result of
a giant impact (e.g., \citealp{Slattery1992}), or a combination of
both (\citealp{MORBIDELLI2012737}). 

Neptune has substantially fewer and mostly smaller satellites than
the other gas planets. The one massive satellite, Triton, is highly
inclined, therefore it was likely captured from a separated KBO binary\textit{
}(\citealp{Agnor_2006}). If Neptune had a primordial (pre-Triton)
satellite system with a mass ratio of $m_{{\rm sat}}/M_{{\rm Nep}}\sim10^{-4}$
as suggested by \citet{Canup_2006}, then Triton\textquoteright s
mass approaches the minimum value required for a retrograde object
to have destroyed the satellite system. Thus, the existence of Triton
places an upper limit on the total mass of such a primordial system. 

The high initial eccentricity of Triton\textquoteright s orbit may
decay by tidal dissipation in less than $10^{9}$ years (\citealp{goldreich1989neptune,Nogueira2011113}).
However, the perturbations from an eccentric Triton destabilize small
irregular satellites (Nereid-like) on a timescale of $10^{5}\ {\rm {\rm y}r}$
(\citealp{Nogueira2011113}). Moreover, \citet{cuk2005constraints}
argue that Kozai cycles increase Triton's mean pericenter, increasing
the circularization timescale beyond the age of the Solar System.
That study proposes that perturbations on pre-existing prograde satellites
induced by Triton lead to mutual disruptive collisions between the
pre-existing satellites. The resulting debris disk interacts with
Triton and drains angular momentum from its orbit, reducing the circularization
timescale to less than $10^{5}$ yr. Yet, it is unclear whether Triton
can induce mutual collisions among such satellites before it experiences
a disruptive collision. Due to its retrograde orbit, collisions between
Triton and a prograde moon would generally have higher relative velocities
than those between two prograde moons. A disruptive collision onto
Triton is inconsistent with its current inclined orbit (\citealp{TritonNereidOrbits}),
as Triton would tend to re-accrete in the local Laplace plane. 

The objective of this study is to explore how interactions (scattering
or collisions) between Triton and putative prior satellites would
have modified Triton\textquoteright s orbit and mass. We evaluate
whether the collisions among the primordial satellites are disruptive
enough to create a debris disk that would accelerate Triton's circularization,
or whether Triton would experience a disrupting impact first. We seek
to find the mass of the primordial satellite system that would yield
the current architecture of the Neptunian system. If the prior satellite
system is required to have a substantially different mass ratio compared
to Jupiter, Saturn and Uranus, this would weaken the hypothesis that
all the satellites in these systems accrete in a similar way. Alternatively,
more stochastic events (e.g. giant impacts) may have a greater influence
on satellite formation for icy giants.

\section{Methods}

We perform N-body integrations using SyMBA code (\citealp{Duncan1998},
based on previous work of \citealp{Wisdom1991}) of a newly captured
Triton together with a hypothetical prograde satellite system for
$10^{7}\ {\rm yr}$ including effects of Neptune\textquoteright s
oblateness. The SyMBA code can effectively resolve close encounters
among orbiting bodies and perfect merger is assumed when an impact
is detected. 

We consider a primordial prograde satellite system composed of 4 satellites
with similar mass ratios compared to Neptune as Ariel, Umbriel, Titania
and Oberon to Uranus \citep{UsatOrbits}. The total mass ratio of
the satellite system is $1.04\times10^{-4}$ (hereafter $M_{{\rm USats}}$),
in agreement with the common mass ratio for gaseous planets predicted
by \citet{Canup_2006}. Triton\textquoteright s initial orbits are
chosen from previous studies of typical captured orbits (\citealp{Nogueira2011113};
see Table \ref{Table Initial Cond} for full list of initial conditions).
We choose to test three retrograde initial inclinations ($105^{o}$,
$157^{o}$, $175^{o}$). Tidal evolution over the simulated time is
small and thus neglected \citep{Nogueira2011113}. For each set of
initial conditions, 10 simulations were performed with randomly varying
longitude of ascending nodes, argument of perihelion and mean anomaly
of all the simulated bodies. In 16 sets of initial conditions the
assumed primordial satellites have the same ratio between the semimajor
axis and planet's Hill sphere as Uranus's satellites. In 4 sets of
initial conditions the exact semimajor axes of Uranus's satellites
are assumed. Using the same initial orbital parameters, we perform
additional simulations with two different satellite system total mass
ratios, $0.35\times10^{-4}$ and $3.13\times10^{-4}$, corresponding
to $0.3\,M_{{\rm USats}}$ and $3\,M_{{\rm USats}}$, respectively.
Overall, our statistics include 200 simulations for each satellite
mass ratio.

\subsection{Disruption Analysis}

We use disruption scaling laws, derived by \citet{Movshovitz201685}
for non-hit-and-run impacts between two gravity-dominated bodies,
to estimate whether the impacts recorded by the N-body code are disruptive.
The scaling laws identify impacts that would disperse half or more
of the total colliding material, regarded hereafter as disruptive
collisions. For head-on impacts, the minimum required kinetic energy
($K^{*}$) to disrupt the target has a linear relation with the gravitational
binding energy of the colliding bodies ($U$) \citep{Movshovitz201685},

\begin{equation}
K^{*}=c_{0}U
\end{equation}

where $c_{0}$ is the slope derived for head-on impacts ($c_{0}=5.5\pm2.9$).

Higher impact angles require higher energies to disrupt a body, since
the velocity is not tangential to the normal plane. A modified impact
kinetic energy is required to incorporate the geometric effects,

\begin{equation}
K_{\alpha}^{*}=\left(\frac{\alpha M_{1}+M_{2}}{M_{1}+M_{2}}\right)K^{*}
\end{equation}

where $\alpha$ is the volume fraction of the smaller body ($M_{2}$)
that intersects the second body ($M_{1}$) \citep{Movshovitz201685,Leinhardt}.
The disrupting relation transforms to:

\begin{equation}
K_{\alpha}^{*}=cU\label{eq:Energy Disruptive}
\end{equation}

where $c$ is the geometrical factor derived from the collision outcomes
with three tested angles. The collisions were tested on a limited
number of impacting angles (direction of velocity relative to the
line connecting the centers at contact, $0^{o}$, $30^{o}$ and $45^{o}$),
therefore, we assume that the relation of $c$ on the impact angle
($\theta$) is given by the following step function:

\begin{equation}
c=c_{_{0}}\left\{ \begin{array}{cc}
1, & \theta<30^{o},\\
2, & 30^{o}\leq\theta<45^{o}\\
3.5, & \theta\geq45^{o}
\end{array}\right..
\end{equation}

It should be noted that ejected material from a satellite collision
can escape the gravitational well of the colliding objects if it has
enough velocity to expand beyond the mutual Hill sphere. For the typical
impacts observed, the required velocity to reach the Hill sphere is
$\sim0.9V_{{\rm esc}}$, where $V_{{\rm esc}}$ is the two-body escape
velocity. The binding energy used in equation (\ref{eq:Energy Disruptive})
to determine the disruption scales as $\sim V_{{\rm esc}}^{2}$, therefore
the required energy to disrupt orbiting bodies may be reduced $\sim0.8K_{\alpha}^{*}$.
Hence, the disruptive scaling laws used may somewhat overestimate
the disrupting energy required, but not too substantially.

\section{Results}

\subsection{Dynamical Survival}

In the 200 simulations with the Uranian satellite system mass ratio
($M_{{\rm USats}}\equiv1.04\times10^{-4}$), we find the overall likelihood
of Triton's survival after $10^{7}\ {\rm yr}$ is $\sim40\%$. Different
sets of initial conditions have different probabilities for Triton\textquoteright s
loss either by escaping the system or falling onto Neptune. For example,
a highly inclined initial Triton ($175{^\circ}$) does not survive
more than $10^{4}\ {\rm yr}$, due to the near alignment of its orbit
with Neptune\textquoteright s equatorial plane which contains the
prograde satellites. In this case, after a final Triton-satellite
collision, the orbital angular momentum of the merged pair is small,
leading to collapse onto Neptune. However, with a lighter satellite
system ($0.3M_{{\rm USats}}$), the post-impact angular momentum is
high enough and Triton survives. In the cases in which Triton survives,
it is usually the last survivor (or one of two remaining satellites
if Triton's initial pericenter is large), reproducing the low number
of Neptunian satellites. Overall, Triton's dynamical survivability
decreases with increasing mass of pre-existing satellite system, $12\%$
for the heavier satellite system with $3\,M_{{\rm USats}}$, and $88\%$
for the lighter satellite system with $0.3\,M_{{\rm USats}}$.

\begin{figure}
\begin{raggedright}
\hspace{0\textwidth}a)\hspace{0.48\textwidth}b)
\par\end{raggedright}
\begin{centering}
\includegraphics[width=0.95\textwidth]{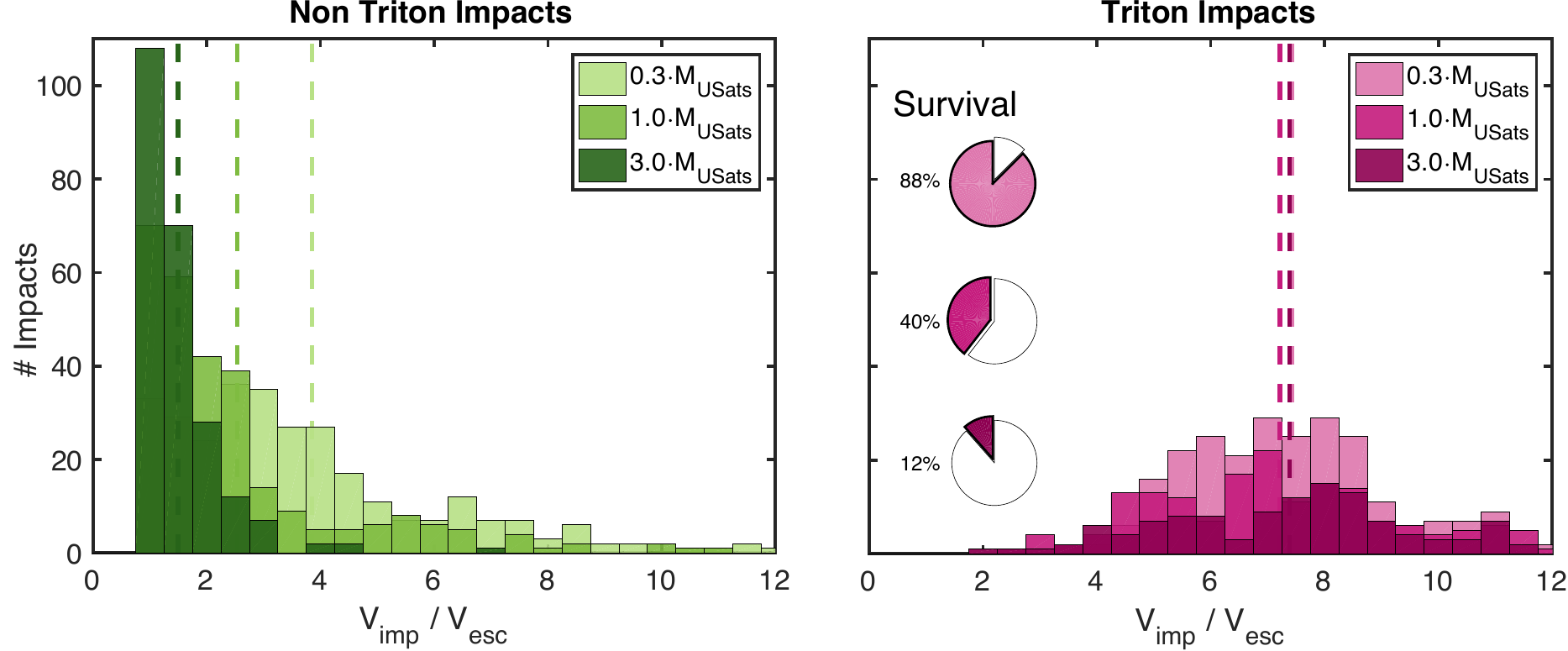}
\par\end{centering}
\caption{\label{fig: Impact Velocity Hist}\textbf{Distribution of impact velocities
between the primordial satellites (left) and onto Triton (right).}
The velocities are normalized by the mutual escape velocity of the
impacting bodies. The dashed lines represent the mean of the distributions
for each assumed primordial system mass (increasing mass represented
by a darker color).}
\end{figure}

Out of the surviving cases, Triton usually ($\sim90\%$) experienced
at least one impact. Due to Triton's initial retrograde orbit, mutual
impacts between the primordial satellites (Figure \ref{fig: Impact Velocity Hist}.a)
have a significant lower velocity compared to the collisions between
Triton and a primordial satellite (Figure \ref{fig: Impact Velocity Hist}.b).
The mean velocity for non-Triton impacts decreases with increasing
mass of pre-existing satellite system, as Triton is less able to excite
the more massive system. 

It should be noted that the number of Triton impacts decreases with
increasing satellite mass, because Triton is lost earlier as the mass
pre-existing satellites increases, and therefore, fewer events are
recorded.

\subsection{Disrupting Impacts}

\begin{figure}
\begin{centering}
\includegraphics[width=0.5\textwidth]{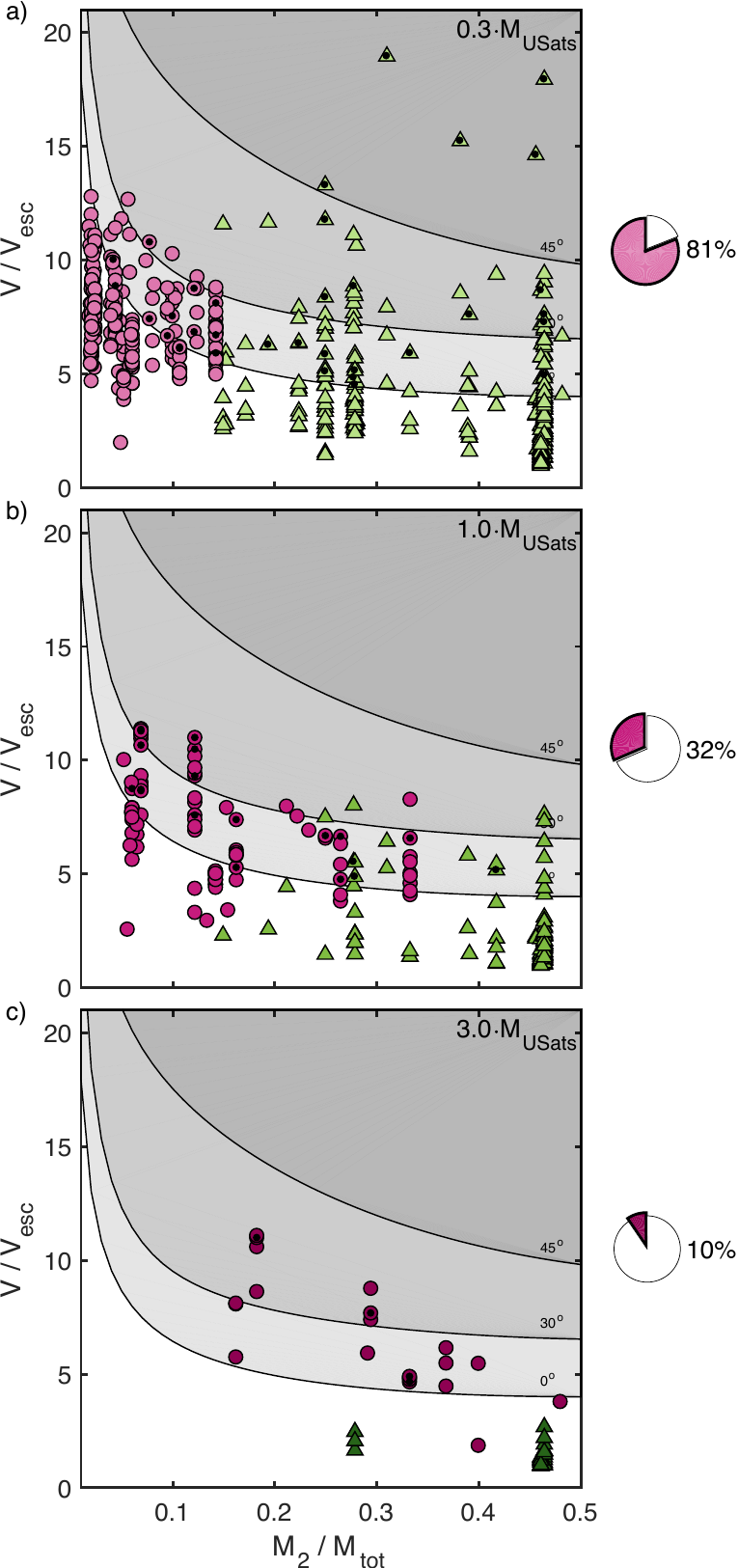}
\par\end{centering}
\caption{\textbf{\label{fig:Impact-phase-space}Impact phase space for all
Triton surviving cases with different primordial satellite mass. }Mass
ratio between the smallest body and the total impacting mass as a
function of normalized impact velocity for impact onto Triton (pink
circles) and between the primordial satellites (green triangles).
The black dots represent disrupting impacts calculated using Movshovitz's
disruption laws \citep{Movshovitz201685}. For simplicity, the disruptive
phase space for $0^{o}$, $30^{o}$ and $45^{o}$ are added as gray
areas. In this way, all impacts on the bottom left of the figure are
non-disruptive and darker areas represent higher disruption probability.
The pie chart on the right indicates the percentage of cases where
Triton dynamically survived and did not experienced any disrupting
impact.}
\end{figure}

For satellite systems with a Uranian mass ratio, impacts onto Triton
are more disruptive ($18\%$ of impacts; pink circles with black dots.
Figure \ref{fig:Impact-phase-space}.b) than mutual collisions among
the primordial satellites ($3\%$; green triangles with black dots,
Figure \ref{fig:Impact-phase-space}.b). With decreasing satellite
mass, disruption of Triton is inhibited as the mass ratio decreases.
Overall, $33\%$ of tested cases with $M_{{\rm USats}}$ ($81\%$
with $0.3M_{{\rm USats}}$ and $10\%$ with $3M_{{\rm USats}}$) resulted
in a stable Triton that did not encounter any disrupting impacts throughout
the evolution. Although Triton does experience disruption in some
cases, a satellite system of $0.3-1\,M_{{\rm USats}}$ has a substantial
likelihood for Triton's survival without the loss of Triton's initial
inclination by disruption and reaccretion into the Laplace plane.

With decreasing satellite mass, disruption for non-Triton bodies increases
somewhat as the typical impact velocity increases ($8\%$ for $0.3M_{{\rm USats}}$
and none for $3M_{{\rm USats}}$). The low rate of primordial satellite
disruption calls into question the formation of a debris disk from
the primordial satellites envisioned by \citealp{cuk2005constraints}
to rapidly circularize Triton. Moreover, assuming that a disruptive
impact between the primordial satellites leads to the formation of
a debris disk, the rubble will quickly settle onto the equatorial
plane and reaccrete to form a new satellite. The timescale of reaccretion
can be estimated by the geometric mass accumulation rate \citep{banfield1992dynamical}
: 

\begin{equation}
\tau_{acc}\sim\frac{m}{\pi r^{2}\sigma_{s}\Omega}\label{eq:5}
\end{equation}

where $m$ is the mass of the satellite, $r$ its radius, $\sigma_{s}$
the surface density of the debris disk and $\Omega$ its orbital frequency.
For typical debris ejection velocities $\sim V_{{\rm esc}}$ (e.g.,\textit{
}\citealp{Benz1999}), the width of the debris ring created by a disrupting
impact is estimated by \textbf{$\Delta a\sim ae\sim a\frac{V_{{\rm esc}}}{V_{{\rm orb}}}$,}
where $a$ is the semimajor axis of the debris, $e$ is the eccentricity
of the debris, $V_{{\rm orb}}$ is the orbital velocity, and $\frac{V_{{\rm esc}}}{V_{{\rm orb}}}$
is proportional to the eccentricity of the debris induced by the impact.
Rearranging the equation above for a disrupted satellite of mass $m$
dispersed across an area of $2\pi a\Delta a$, we obtain:

\begin{equation}
\tau_{acc}\sim\frac{2a^{2}}{r^{2}\Omega}\cdot\frac{V_{{\rm esc}}}{V_{{\rm orb}}}
\end{equation}

Assuming an impact between two primordial satellites equivalent to
the mass of Oberon and Titania at a distance of $10R_{{\rm Nep}}$,
the time scale for reaccretion is $\sim10^{2}\ {\rm yr}$. The reaccretion
time scale is smaller than the evaluated Triton's eccentricity decay
time of $10^{4}-10^{5}\ {\rm yr}$ \citep{cuk2005constraints}. Thus,
the debris disk, even if it formed, would likely re-accrete before
Triton's orbit circularized.

\subsection{Final Triton's Orbits}

In order to ensure stabilization of Nereid (and Nereid-like satellites),
Triton's apoapse needs to decrease to within Nereid's orbit in $10^{5}\ {\rm yr}$
\citep{cuk2005constraints,Nogueira2011113}. Otherwise Neptune's irregular
satellites must be formed by an additional subsequent process (e.g.,
\citealp{Nogueira2011113}). Torques produced by the Sun and Neptune's
shape are misaligned by Neptune's obliquity ($30^{o}$), causing a
precession of the argument of pericenter (Kozai oscillations). For
large orbits, the Kozai mechanism induces oscillations (period $\sim10^{3}\ {\rm yr}$)
in the eccentricity and inclination such that the z-component of the
angular momentum ($H_{z}=\sqrt{1-e^{2}}\cos I$, where $I$ is the
Triton's inclination with the respect to the Sun) is constant. For
retrograde orbits, the eccentricity and inclination oscillate in phase,
therefore the maximum inclination occurs at the maximum eccentricity.
For small orbits ($<70R_{{\rm Nep}}$; \citealp{Nogueira2011113}),
the torque induced by Neptune's shape is larger than the Kozai cycles
induced by the Sun, and Triton's tidal decay occurs with approximately
constant inclination. Here we seek to identify Triton analogs with
final apoapses smaller than Nereid's pericenter ($55\,R_{{\rm Nep}}$;
\citealp{TritonNereidOrbits}) and inclinations close to Triton's
current inclination of $157^{o}$, as this will remain constant in
subsequent tidal circularization. 

\begin{figure}
\begin{centering}
\includegraphics[width=0.5\textwidth]{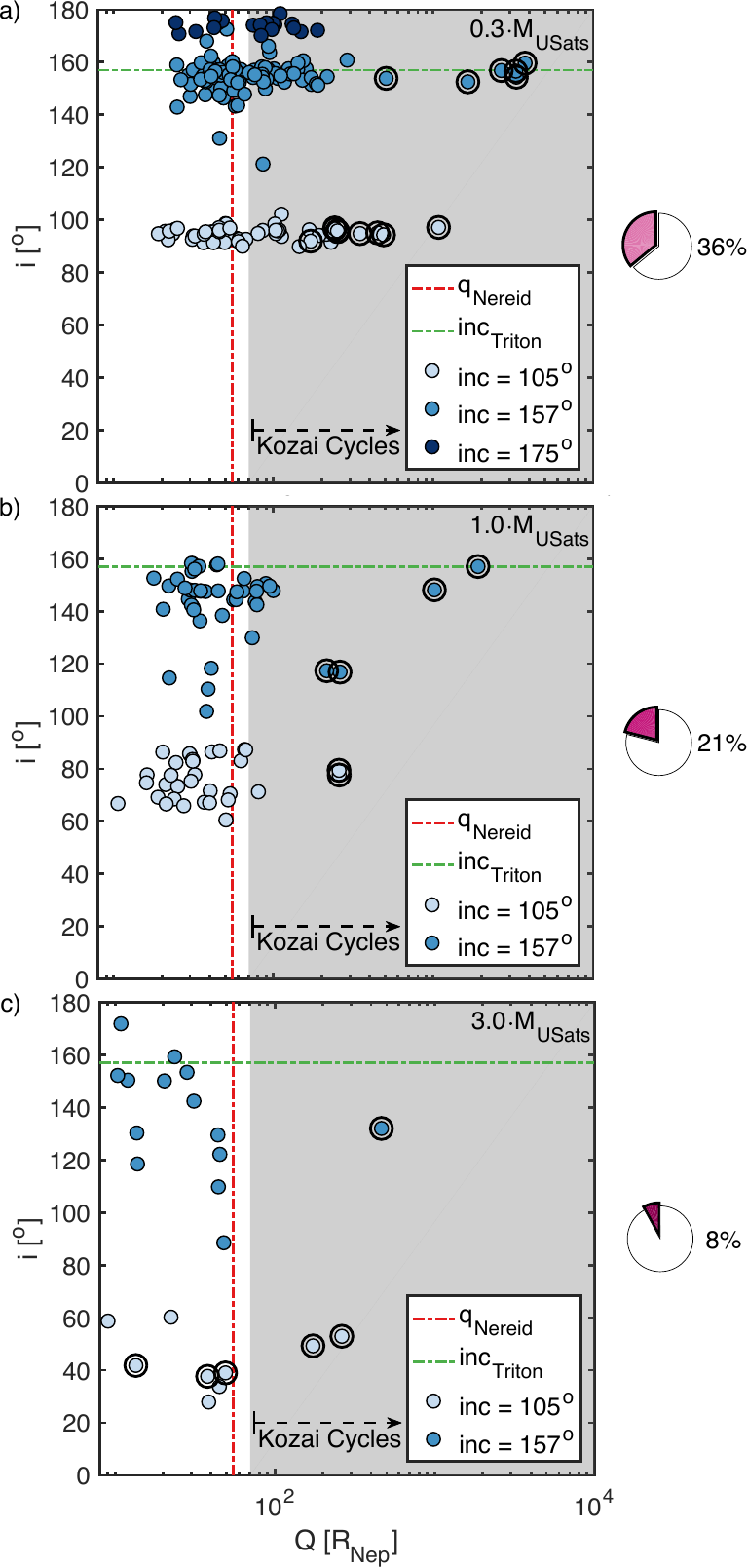}
\par\end{centering}
\caption{\label{fig:Triton=002019s-final-apoapsis}\textbf{Final Triton orbits}.
Triton\textquoteright s final apoapsis in Neptune Radii vs. its final
orbital inclination for different initial inclinations. The vertical/horizontal
dashed line represents the current Nereid\textquoteright s periapsis/Triton\textquoteright s
current inclination. The light gray region represents the regions
where Kozai perturbations are significant; for lower orbits, tidal
evolution proceeds with constant inclination (\citealp{Nogueira2011113}).
Simulated Triton analogs that did not experience any impacts are indicated
by the black circles. The pie charts on the right indicate the percentage
of cases where Triton dynamically survived, did not experienced any
disrupting impact, and has a final orbit within Nereid's current orbit.}

\end{figure}

As seen previously, the smallest pre-existing satellite system mass
ratio has a high rate of survival, and about half of the Triton's
analogs are within Nereid's orbit with minimal inclination change
(Figure \ref{fig:Triton=002019s-final-apoapsis}.a). As the primordial
satellites mass increases, the percentage of Triton analogs inside
Nereid's orbit increases, although Triton's inclination change is
larger. Moreover, due to the small survivability rate, only a small
number of cases with $3M_{{\rm USats}}$ fulfilled all the required
conditions to be regarded as successful Triton analogs that may allow
Nereid-like satellites to survive (Triton's dynamically survival with
no disruption, on a small and inclined orbit). Typically, Triton analogs
that did not experience any impact have larger final orbits, as scattering
alone does not effectively decrease the orbital apoapse. 

In addition, we roughly estimated the effect of solar perturbations
on Triton's orbital evolution by performing additional simulations
that include Triton, Neptune and the Sun (positioned on an inclined
orbit relative to Neptune's equatorial plane in order to mimic the
planet's obliquity; see Appendix B for more details). Due to induced
eccentricity oscillations, Triton spends only $\sim10\%$ of the time
in the region populated by primordial satellites (\citealp{cuk2005constraints},
Figure 1). Out of the recognized successful cases in Figure \ref{fig:Triton=002019s-final-apoapsis},
$\sim30-50\%$ are already within Nereid's orbit after $10^{4}\ {\rm yr}$
(see Figure \ref{fig:Triton=002019s-apoapsis-in}). Even though Kozai
could lengthen the time of orbit contraction by a factor of $10$,
Triton's orbit in these cases would still decrease within Nereid's
orbit in $\leq10^{5}\ {\rm yr}$. Therefore, even if Solar perturbations
were included in the numerical scheme, Nereid would still likely be
stable in these cases. Additional studies are needed to determine
the specific details induced by the Kozai mechanism in the first part
of Triton's evolution when the semimajor axis is still large to fully
evaluate the stability of pre-existing irregular moons including the
effects of a pre-existing prograde satellite system.

\section{Discussion}

We performed dynamical analyses of a newly captured Triton together
with a likely primordial Neptunian satellite system. Most of the recorded
impacts (onto Triton or between the primordial satellites) are not
disruptive, therefore we conclude that the formation of a debris disk
composed of primordial satellite material is unlikely for the assumed
initial conditions. Moreover, if a debris disk is indeed formed outside
the Roche limit, its reaccretion time scale is $\sim10^{2}\ {\rm yr}$,
smaller than Triton's orbital decay by the debris disk, which is $10^{4}-10^{5}\ {\rm yr}$
\citep{cuk2005constraints}. 

We find that a primordial satellite system of $>0.3\,M_{{\rm USats}}$
decreases Triton's orbit within Nereid's via collisions and close
encounters. Triton's interactions with the primordial system enhances
its circularization and may preserve the small irregular satellites
(Nereid-like), that might otherwise be lost during a protracted Triton
circularization via tides alone, echoing \citet{cuk2005constraints}
findings although through a different mechanism. The Kozai mechanism
may prolong the timing of such impacts, however we found cases where
Triton's circularization still appears fast enough for Nereid's stability.

Moreover, we find that a primordial satellite system of $0.3-1\,M_{{\rm USats}}$
has a substantial likelihood for Triton's survival while still maintaining
an initial high inclination. Higher mass systems have a low probability
of reproducing the current system ($\leq10\%$). We conclude that
a primordial satellite system of a mass ratio comparable to that of
Uranus's current system appears consistent with the current Neptunian
system and offers a means to potentially preserve pre-existing irregular
Nereid-like satellites.

\acknowledgements{We thank Oded Aharonson for valuable comments and suggestions and
Julien Salmon for providing guidance on the computational code. This
project was supported by the Helen Kimmel Center for Planetary Science,
the Minerva Center for Life Under Extreme Planetary Conditions, and
by the I-CORE Program of the PBC and ISF (Center No. 1829/12). R.R.
is grateful to the Israel Ministry of Science, Technology and Space
for their Shulamit Aloni fellowship. R.M.C. was supported by NASA's
Planetary Geology and Geophysics program.}

\software{SyMBA \textit{\citep{Duncan1998}}}

\appendix

\begin{table}
\begin{centering}
\begin{tabular}{|c|c|c|c|c|c|c|c|c|c|c|c|c|}
\hline 
\multirow{2}{*}{{\small{}\# Set}} & \multirow{2}{*}{{\small{}$a_{T}[R_{Nep}]$}} & \multirow{2}{*}{{\small{}$q_{T}[R_{Nep}]$}} & \multirow{2}{*}{{\small{}$inc_{T}[deg]$}} & \multicolumn{3}{c|}{\#Triton's Survival} & \multicolumn{3}{c|}{\# Triton's Fall onto Planet} & \multicolumn{3}{c|}{\#Triton's Escape}\tabularnewline
\cline{5-13} 
 &  &  &  & {\scriptsize{}$0.3M_{{\rm USats}}$} & {\scriptsize{}$M_{{\rm USats}}$} & {\scriptsize{}$3M_{{\rm USats}}$} & {\scriptsize{}$0.3M_{{\rm USats}}$} & {\scriptsize{}$M_{{\rm USats}}$} & {\scriptsize{}$3M_{{\rm USats}}$} & {\scriptsize{}$0.3M_{{\rm USats}}$} & {\scriptsize{}$M_{{\rm USats}}$} & {\scriptsize{}$3M_{{\rm USats}}$}\tabularnewline
\hline 
\hline 
1$^{*}$ & 300 & 8.1 & 105 & 9 & 8 & 2 & 0 & 1 & 6 & 1 & 1 & 2\tabularnewline
\hline 
2$^{*}$ & 300 & 8.1 & 157 & 10 & 7 & 3 & 0 & 3 & 7 & 0 & 0 & 0\tabularnewline
\hline 
3 & 300 & 8.1 & 157 & 10 & 0 & 1 & 0 & 10 & 6 & 0 & 0 & 3\tabularnewline
\hline 
4 & 1004 & 8.0 & 157 & 9 & 0 & 0 & 0 & 8 & 7 & 1 & 2 & 3\tabularnewline
\hline 
5 & 1004 & 8.0 & 105 & 6 & 4 & 0 & 1 & 2 & 4 & 3 & 4 & 6\tabularnewline
\hline 
6 & 128 & 8.0 & 157 & 10 & 2 & 0 & 0 & 8 & 10 & 0 & 0 & 0\tabularnewline
\hline 
7 & 128 & 8.0 & 105 & 10 & 5 & 2 & 0 & 5 & 7 & 0 & 0 & 1\tabularnewline
\hline 
8 & 512 & 8.0 & 157 & 9 & 1 & 1 & 0 & 7 & 5 & 1 & 2 & 4\tabularnewline
\hline 
9 & 512 & 8.0 & 105 & 8 & 4 & 1 & 0 & 2 & 5 & 2 & 4 & 4\tabularnewline
\hline 
10 & 2000 & 8.0 & 157 & 1 & 1 & 0 & 0 & 2 & 1 & 9 & 7 & 9\tabularnewline
\hline 
11 & 2000 & 8.0 & 105 & 2 & 2 & 1 & 0 & 2 & 1 & 8 & 6 & 8\tabularnewline
\hline 
12 & 300 & 6.8 & 105 & 10 & 3 & 2 & 0 & 2 & 5 & 0 & 5 & 3\tabularnewline
\hline 
13 & 512 & 8.0 & 175 & 10 & 0 & 0 & 0 & 9 & 9 & 0 & 1 & 1\tabularnewline
\hline 
14 & 300 & 6.8 & 175 & 10 & 0 & 0 & 0 & 10 & 8 & 0 & 0 & 2\tabularnewline
\hline 
15$^{*}$ & 512 & 8.0 & 157 & 10 & 7 & 0 & 0 & 2 & 9 & 0 & 1 & 1\tabularnewline
\hline 
16$^{*}$ & 512 & 8.0 & 105 & 10 & 7 & 1 & 0 & 0 & 3 & 0 & 3 & 6\tabularnewline
\hline 
17 & 512 & 12.5 & 157 & 10 & 4 & 2 & 0 & 5 & 5 & 0 & 1 & 3\tabularnewline
\hline 
18 & 512 & 17.5 & 157 & 10 & 6 & 1 & 0 & 2 & 4 & 0 & 2 & 5\tabularnewline
\hline 
19 & 1004 & 29 & 157 & 10 & 9 & 1 & 0 & 0 & 5 & 0 & 1 & 4\tabularnewline
\hline 
20 & 1004 & 37.9 & 157 & 10 & 9 & 5 & 0 & 0 & 0 & 0 & 1 & 5\tabularnewline
\hline 
\end{tabular}
\par\end{centering}
\caption{\label{tab:Table Initial Cond}Survival Results - $a_{T}$, Triton's
initial semimajor axis in Neptune Radii, $q_{T}$, Triton's initial
pericenter in Neptune Radii, $inc_{T}$, Triton's initial inclination.
$^{*}$Semimajor axis are similar to Ariel, Umbriel, Titania and Oberon.}
\end{table}

\section{Timing of Impacts}

\begin{figure}[H]
\begin{raggedright}
\hspace{0.07\textwidth}a)\hspace{0.4\textwidth}b)
\par\end{raggedright}
\begin{centering}
\includegraphics[width=0.4\textwidth]{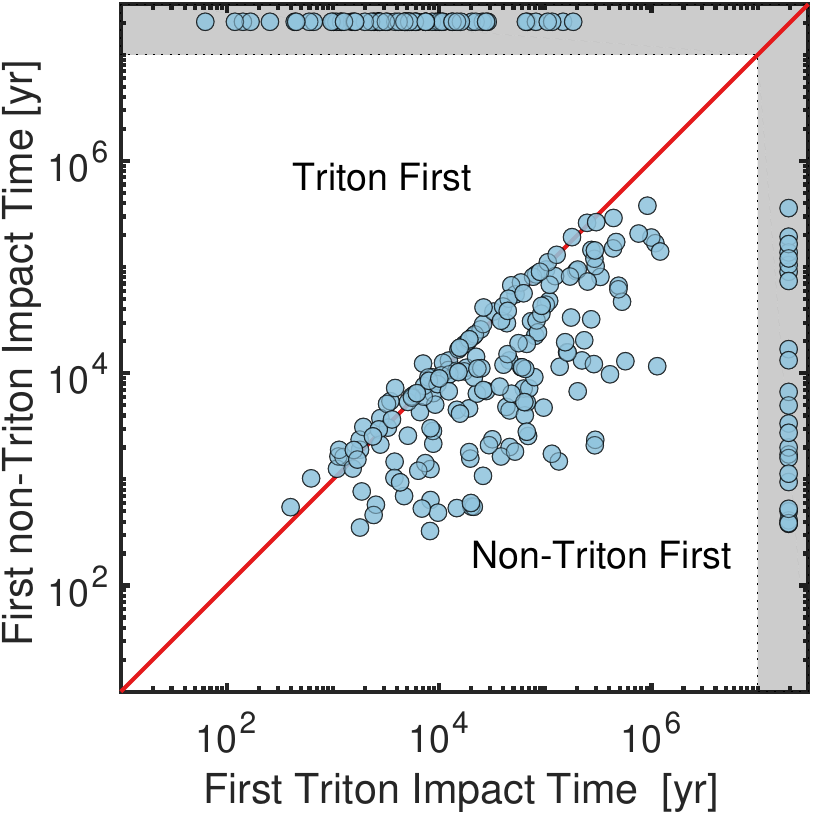}\hspace{0.03\textwidth}\includegraphics[width=0.4\textwidth]{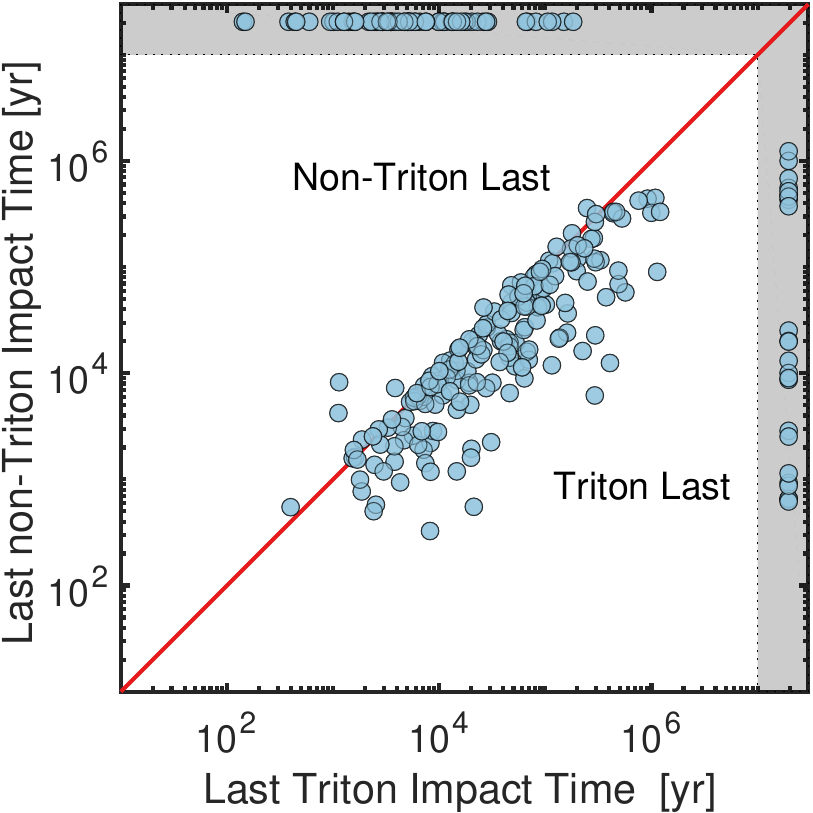}
\par\end{centering}
\caption{\label{fig:Timing-of-first}a) The first impact onto Triton vs the
first impact between primordial satellites for the Triton-surviving
cases. Markers that are above/bellow the red line represent cases
where Triton/primordial satellites experience the first impact. b)
The last impact onto Triton vs the first impact between primordial
satellites for the Triton-surviving cases. Markers that are above/bellow
the red line represent cases where primordial satellites/Triton experience
the first/last impact. Cases where primordial satellites/Triton did
not encounter any impact are represented by the horizontal/vertical
grey area. }
\end{figure}

The bodies start colliding after $\sim10^{2}-10^{3}\ {\rm yr}$ (Figure
\ref{fig:Timing-of-first}.a). The first recorded impact is usually
among the primordial satellites, consistent with previous estimations
\citep{banfield1992dynamical,cuk2005constraints}. In a small number
of cases, the primordial satellites did not impact themselves (horizontal
markers inside the grey area), but Triton cannibalized the entire
system. Usually a first Triton impact will lead to a primordial impact
soon after (markers that are above but close to the red line). Moreover,
Triton will encounter the last impact in most of the scenarios (Figure
\ref{fig:Timing-of-first}.b). All impacts (Figure \ref{fig:Timing-of-first}.b)
occur within the first $\sim10^{6}\ {\rm yr}$ of the simulation while
the difference between most impacts is $10^{3}-10^{4}\ {\rm yr}$.

\section{\label{sec:Solar-Perturbations}Solar Perturbations}

In order to estimate the effect of perturbation induced by the Sun
we performed simulations using a Bulirsch-Stoer integrator to simulate
Triton as a massless particle orbiting Neptune with no satellites.
The sun is added as a secondary orbiting body with an inclined orbit
equal to Neptune's obliquity. We used the same Triton's initial conditions
as used before and checked whether Triton remains in orbit after $10^{7}\ {\rm yr}$.

\begin{figure}
\begin{centering}
\includegraphics[width=0.4\textwidth]{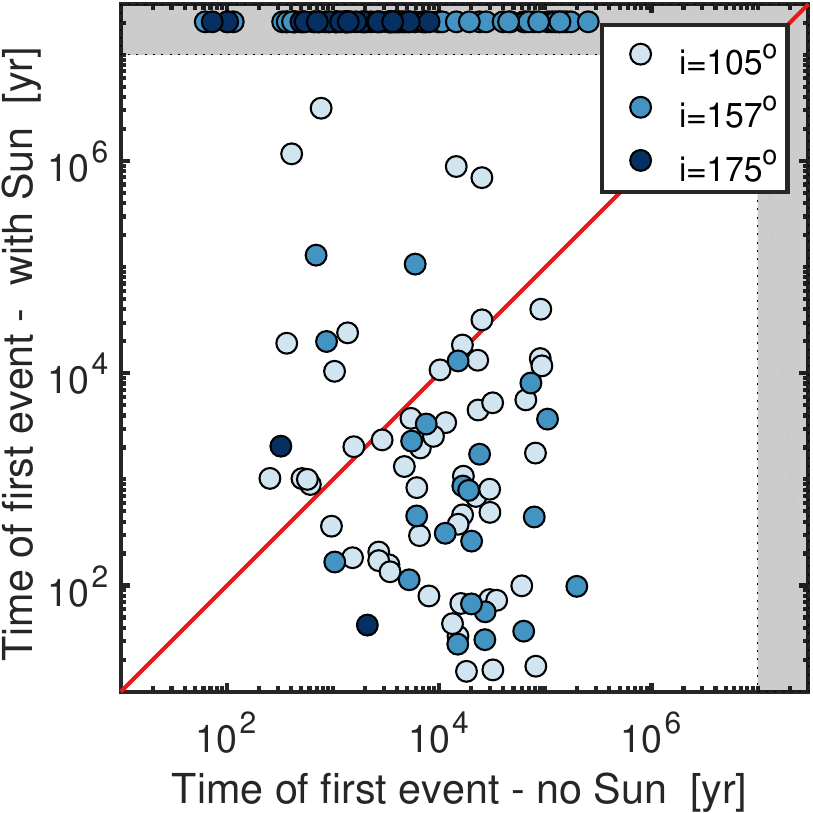}
\par\end{centering}
\caption{\label{fig:Sun Timing}Timing of the first event in the simulation
with a primordial system of $M_{{\rm USats}}$ vs Triton's loss time
due to Kozai cycles. Markers that are bellow the red line represent
cases where loss due to Kozai perturbations is faster than the satellite
dynamics. Cases where Triton survived the Sun perturbation simulations
are represented in the horizontal grey area. }
\end{figure}

For the lower inclined orbits ($105^{o}$), we find that in $88\%$
of the cases Triton fell onto Neptune as the Kozai mechanism is strongest
when the torque is perpendicular ($I\sim90^{o}$) \citep{Nogueira2011113}.
Moreover, the timing of Triton is lost due to Sun perturbations is
usually earlier than the collision timescale with a primordial satellite
system of $M_{{\rm USats}}$ (markers bellow the red line in the white
region of Figure \ref{fig:Sun Timing}). For the higher inclined orbits
($157^{o}$/$175^{o}$) we find significantly lower percentages ($26\%$/$10\%$)
of Triton's loss. Due to eccentricity perturbations induced by the
Sun, only $10\%$ of Triton's orbits will cross the primordial satellite
system \citep{cuk2005constraints}. Therefore, Triton's collisions
with the primordial satellites will be prolonged by roughly a factor
of $10$ relative to previous simulations that do not include solar
perturbations.\textbf{ }

\begin{figure}
\begin{centering}
\includegraphics[scale=0.5]{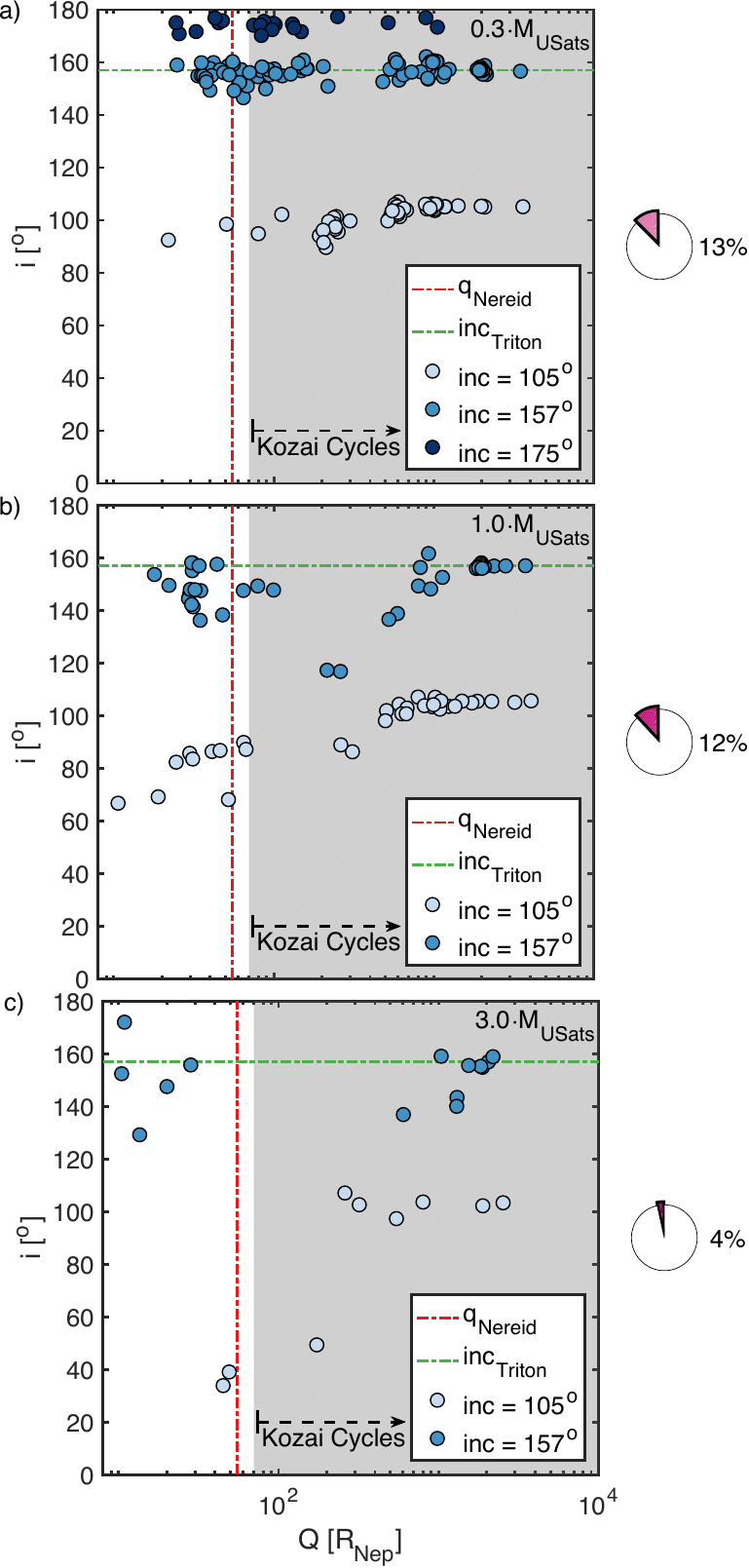}
\par\end{centering}
\caption{\label{fig:Triton=002019s-apoapsis-in}Triton\textquoteright s apoapsis
in Neptune Radii vs. its orbital inclination for different initial
inclinations at $10^{4}\ {\rm yr}$. The vertical/horizontal dashed
line represents the current Nereid\textquoteright s periapsis/Triton\textquoteright s
current inclination. The light gray region represents the regions
where Kozai perturbations are significant. The pie charts on the right
indicate the percentage of cases where Triton dynamically survived
and has an orbit within Nereid's current orbit after $10^{4}\ {\rm yr}$
of simulated time.}
\end{figure}

Figure \ref{fig:Triton=002019s-apoapsis-in} shows the momentary Triton's
orbit at $10^{4}\ {\rm yr}$ after the start of the simulation. We
find that $\sim30-50\%$ of successful cases in Figure \ref{fig:Triton=002019s-final-apoapsis}
are already within Nereid's orbit at this time. In these cases, Triton's
prolonged evolution should still be consistent with Nereid's presence
in its current orbit even considering Kozai oscillations. 

\bibliographystyle{aasjournal}
\bibliography{TritonBib}

\end{document}